\documentclass[twocolumn,twoside,10pt,aps,prl,superscriptaddress,amsmath,amssymb,longbibliography]{revtex4-2}
\usepackage[english]{babel}
\usepackage{lineno}
\usepackage{amssymb}
\usepackage{dcolumn}
\usepackage{bm}
\usepackage{graphicx}
\usepackage{amsmath}
\usepackage{color}
\usepackage{orcidlink}
\hypersetup{hidelinks} 
\usepackage{hyperref}

\def\bra#1{\left\langle{#1}\right|}
\def\ket#1{\left|{#1}\right\rangle}
\def\braket#1#2{\left\langle{{#1}}\mathrel{\left|{\vphantom{{#1}{#2}}}\right.\kern-\nulldelimiterspace}{{#2}}\right\rangle}

\begin{document}
\title{Quantum double lock-in detection via sequential orthogonal quantum mixing}
\author{Min Zhuang}\thanks{These authors contributed equally}
\affiliation{Institute of Quantum Precision Measurement, State Key Laboratory of Radio Frequency Heterogeneous Integration, College of Physics and Optoelectronic Engineering, Shenzhen University, Shenzhen 518060, China}
\author{Sijie Chen}\thanks{These authors contributed equally}
\affiliation{Laboratory of Quantum Engineering and Quantum Metrology, School of Physics and Astronomy, Sun Yat-Sen University (Zhuhai Campus), Zhuhai 519082, China}
\author{Jiahao Huang}
\affiliation{Laboratory of Quantum Engineering and Quantum Metrology, School of Physics and Astronomy, Sun Yat-Sen University (Zhuhai Campus), Zhuhai 519082, China}
\author{Chaohong Lee}
\altaffiliation{Email: chleecn@szu.edu.cn, chleecn@gmail.com}
\affiliation{Institute of Quantum Precision Measurement, State Key Laboratory of Radio Frequency Heterogeneous Integration, College of Physics and Optoelectronic Engineering, Shenzhen University, Shenzhen 518060, China}
\affiliation{Quantum Science Center of Guangdong-Hong Kong-Macao Greater Bay Area (Guangdong), Shenzhen 518045, China}
    

\date{\today}
		
\begin{abstract}
High-precision measurement of oscillating signal is a ubiquitous issue in fundamental science and a critical task in practical technologies.  
In quantum metrology, quantum lock-in detection provide an efficient method for measuring such signal.
In general, when the initial phase of the oscillating signal is unknown, quantum double lock-in detection can effectively extract complete information about the signal's amplitude, frequency, and initial phase. 
Conventional quantum double lock-in detection requires two individual quantum interferometry, each of which must undergo state preparation and readout.
However, the time for state preparation and readout need not be negligible in practical experiments. 
In particular, the time for state preparation is longer than the time for
sensing.
To save experimental resources, it is challenging to achieve quantum double lock-in detection just via a single quantum interferometry while still extracting complete information about the signal’s amplitude, frequency, and initial phase.
Here, we present a general protocol for achieving a quantum double lock-in detection just via a single quantum interferometry under a sequential orthogonal periodic multipulse sequences.
In particular, if the input state is a Greenberger-Horne-Zeilinger state and two interaction-based operations are applied during interferometry, the measurement precisions for frequency, amplitude, and initial phase can both approach the Heisenberg limit. 
Our study paves a new way for measuring oscillating signals with a single quantum interferometry, and provides a feasible method for achieving Heisenberg-limited detection of alternating signals.
\end{abstract}	
\maketitle

\section{Introduction\label{Sec1}}
High-precision measurement of oscillating signal is an ubiquitous issue in fundamental science and a critical task in practical technologies~\cite{Helstrom,BMEscher2011,RDemkowiczDobrz2012,CLDegen2017,Vengalattore2007,Helstrom1967,Paris2009}. 
Quantum lock-in detection is an efficient method of measuring an oscillating signal with known initial phase detection~\cite{Nature473,PRXQUANTUM2040317,MZhuang2024,NatCommun814157}.
Further, when the initial phase of the oscillating signal is unknown, one can use two individual quantum lock-in detection to realize a double quantum lock-in detection to extract the complete characteristics of the target signal~\cite{Chen2024}.
The conventional quantum lock-in detection  requires two individual quantum systems to realize two individual quantum interferometry, each of which must undergo state preparation and
readout.
However, the time for state preparation and readout need not be negligible in practical experiments.
In particular, the time for preparation of entangled state is longer than the time for sensing.
Therefore, to save experimental resources, achieving a quantum double lock-in detection just via a single quantum interferometry remains a challenging endeavor.


It is well known that many-body quantum entanglement is a useful resource to improve the measurement precision over individual particles~\cite{PRL96010401,Science306,NatPhotonics5222,AnnuRevColdAtomsMolecule,PRL97150402,BLu2019,Huang2024,PRL102070401,Science306,Science316,Nature450,NatPhotonics5,Lucke2011,RevModPhys90035005,JJBollinger1996}.
For $N$ individual particles, the measurement precision scales as the standard quantum limit (SQL) (i.e., $\propto 1/\sqrt{N}$).
The SQL can be surpassed by using quantum entanglement.
Especially, by inputting the Greenberger-Horne-Zeilinger (GHZ) states, the measurement precision can be improved to the Heisenberg limit (i.e., $\propto 1/N$)~\cite{PRL96010401,Science306,NatPhotonics5222,AnnuRevColdAtomsMolecule,PRL97150402,BLu2019,Huang2024,JJBollinger1996}.
Meanwhile, many-body quantum entanglement also can be used to increase the measurement precision of a quantum lock-in detection and even achieve the Heisenberg-limited detection of the target signal~~\cite{PRXQUANTUM2040317,MZhuang2024,JZhang2026}.
%
Based on the quantum-enhanced oscillating signal estimation via quantum lock-in detection, it is natural to ask the following:
(i) To save experimental resources, can one achieve a quantum double lock-in detection just via a single quantum interferometry while still extracting the complete information about the signal’s amplitude, frequency, and initial phase?
(ii) Can one use many-body quantum entanglement to make the measurement precision of the target signal surpass the SQL or even attain the Heisenberg-limited scaling?

In this article, we present a general protocol for achieving quantum double lock-in detection just via a single quantum interferometry under sequential orthogonal periodic multipulse sequences.
Our quantum interferometry can be divided into three steps: state preparation, signal interrogation, and readout. 
To extract the complete information of an unknown oscillating signal, we apply two orthogonal periodic multipulse sequences in the signal interrogation stage.
Thus, the interrogation process is divided into two signal-accumulation processes and linked via a unitary operation.
In our scheme, the pulse spacing $\tau_{m}$ is adjusted to probe the oscillating frequency $\omega$ of the target signal. 
For a given evolution time $T$, at lock in point $\tau_{m}=\tau$ with $\tau=\pi/\omega$, the accumulated phase is a constant.
While for $\tau_{m}\neq\tau$, the accumulated phase oscillates dramatically and is sensitively dependent on the $\tau-\tau_m$.
This accumulated phase determines the half-population, which is an exactly double sinusoidal oscillation just at lock in point.
Thus one may infer the lock-in point by measuring the half-population.
Further, the signal amplitude and initial phase can also be extracted.
Moreover,  we find that if the initial state is prepared as a GHZ state, and suitable interaction-based operations are applied in the interrogation and readout stages, the measurement precisions of the frequency, amplitude and initial phase of the oscillating signal can both exhibit Heisenberg-limited scaling.
Compared to the conventional double lock-in detection using two individual quantum
interferometry, our scheme just require a single quantum interferometry and thus reduce experimental resources.
Overall, our proposed protocol not only provides an effective pathway to extract the complete information of an oscillating signal, but also provides a feasible method for achieving Heisenberg-limited detection of oscillating signals.

This article is organized as follows.
In Sec.~\ref{Sec2}, we introduce our general protocol for a many-body quantum double lock-in detection via sequential orthogonal quantum mixing.
In Sec.~\ref{Sec3}, we study the measurement precisions for different input states.
We demonstrate how to achieve the Heisenberg-limited quantum double lock-in detection via GHZ states.
In Sec.~\ref{Sec4}, we discuss the robustness of our scheme.
Finally, we give a brief summary and discussion in Sec.~\ref{Sec5}.
\section{General scheme \label{Sec2}}
\begin{figure*}[!htp]
\includegraphics[width=2\columnwidth]{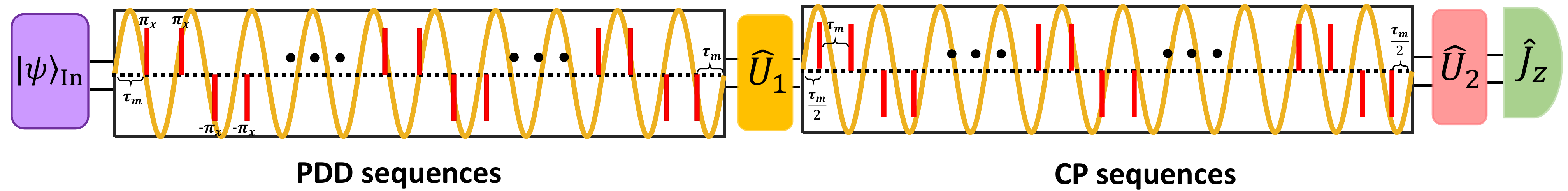}
\caption{\label{Fig1}
%
%
%
%
%
%
%
%
%
The quantum double lock-in detection via sequential orthogonal quantum
mixing.
The scheme includes three stages: (i) initialization, (ii) interrogation, and (iii) readout.
%
In the initialization stage, a suitable probe $\ket{\psi}_\textrm{in}$ is prepared.
Then, the input state undergoes an interrogation stage for signal accumulation.
The interrogation stage is divided into two signal accumulation processes and linked via a beam splitters $\hat{U}_{1}$.
In the first accumulation process, the PDD sequences is applied as the mixing modulations.
In the second accumulation process, the CP sequences is applied as the mixing modulations.
In the readout stage, a certain unitary operation $\hat{U}_{2}$ is applied for recombination, and the half-population difference is measured.
The quantum double lock-in detections can extract the complete characteristics of the target signal $\emph{S}(t)=A \sin(\omega t+\beta)$.
}
\end{figure*}

%
%
To illustrate our scheme, we consider an ensemble of $N$ identical two-state bosonic particles.
Potential probes include Bose condensed atoms~\cite{Science345424,Science355620,PNAS1156381,PRL111143001,PRL113103004}, trapped ions~\cite{NaturePhys5,Science3521297,Science3641163,Science3041476,PRL95060502,TMonz2011}, nitrogen-vacancy centers~\cite{NanoLetters215143,NanoLett208267,NanoLett202980,NaturePhys4810}, doped spins in semiconductors~\cite{PRL107166802} and so on.
The two levels are chosen as two magnetic levels and are labeled as $\ket{\uparrow}$ and $\ket{\downarrow}$ hereafter, respectively.
Introducing the collective spin operators,
$\hat{J}_{x}=\frac{1}{2}(\hat{a}^{\dag}\hat{b}+\hat{a}\hat{b}^{\dag}),\hat{J}_{y}=\frac{1}{2i}(\hat{a}^{\dag}\hat{b}-\hat{a}\hat{b}^{\dag}),
\hat{J}_{z}=\frac{1}{2}(\hat{a}^{\dag}\hat{a}-\hat{b}^{\dag}\hat{b})$, and the probe state can be represented in terms of the Dicke basis $\{|J,m\rangle\}$ with $J=\frac{N}{2}$ and $m = -J,-J + 1, ..., J+1, J$.
Here, $\hat{a}$ and $\hat{b}$ respectively denote annihilation operators for particles in $\ket{\uparrow}$ and $\ket{\downarrow}$.
Given the probe as an ensemble of two-mode bosonic particles, the coupling between the probe and the target signal [an ac magnetic field $B(t) = A \sin(\omega t +\beta)\hat{z}$] is described by the Hamiltonian
\begin{equation}\label{Eq:HamS}
\hat{H}_{\textbf{B}}=\gamma{\textbf{B}} \cdot \textbf{J}=\gamma B \sin(\omega t+\beta) \hat{J}_{z},
\end{equation}
where $B$ corresponds to the magnetic field amplitude, $\gamma$ is the gyromagnetic ratio, $\omega$ denotes the oscillation frequency, and $\beta$ is the initial phase.
For convenience, we assume $\gamma=1$.
In analogy to the classical double lock-in detection, our goal is to measure $A$, $\omega$ and $\beta$.
Here, we consider the external modulation $\hat{H}_\textrm{ref}=\frac{\hbar}{2}\Omega (t)\hat{J}_{x}$ is designed as a sequence of $\pi$ pulses with equidistant spacings, which periodically rotates the sensor around the $\hat{x}$ direction.
Thus the whole Hamiltonian becomes
\begin{equation}\label{H_int}
{\hat{H}}=\gamma B \sin(\omega t+\beta) \hat{J}_{z}+\Omega({t})\hat{J}_{x}.
\end{equation}
%
%
%
%
%
For simplicity, we assume every $\pi$ pulse is sharp enough and its time duration can be neglected.
Thus, the {$\pi$-pulse} sequences can be described by instantaneous pulses
\begin{equation}\label{PDD,CP}
\Omega(t)=\pi\sum_{m=1}^{n}\delta[t-(j-\lambda)\tau_m],
\end{equation}
with $\delta(t)$ being the Dirac $\delta$ function, $n$ denotes the pulse number, $\tau_m$ describing the spacing of the adjacent $\pi$ pulses, and the parameter $\lambda$ determining the relative phase with respect to the target signal.
As expected, conventional quantum double lock-in detection can be realized by combining two individual quantum interferometry with two orthogonal
periodic multi-pulse sequences to extract the complete characteristics of the target signal, such as a PDD sequence ($\lambda=0$) and a CP sequence ($\lambda= 1/2$), as illustrated in Ref.\cite{Chen2024}.
%
%
However, the time for state preparation and readout need not be negligible in practical experiments.
In particular, the time for state preparation is longer than the time for sensing.
Consequently, the realization of a quantum double lock-in detection just via a single quantum interferometry to save experimental resources remains a challenge.

To address this challenge, we present a general protocol for achieving double lock-in detection via sequential orthogonal quantum mixing.
The scheme includes three stages: (i) initialization, (ii) interrogation, and (iii) readout, as shown in Fig~\ref{Fig1}.
%
In the initialization stage, a suitable probe $\ket{\psi}_\textrm{in}$ is prepared.
Then, the input state undergoes an interrogation stage for signal accumulation.
The interrogation stage is divided into two signal accumulation processes and linked via a beam splitters $\hat{U}_{1}$.
For convenience, we assume the time for the two accumulation processes both are $T=n\tau_m$ with $n$ chosen as a even number to suppress DC noise. 
In the first phase accumulation process,  the mixing modulations $H_\textrm{mix1}=\Omega_\textrm{PDD}(t)\hat{J}_{x}$ (implemented by the PDD sequences) is applied to obtain the phase accumulation $\phi^\textrm{PDD}_n$.
In the second phase accumulation process, the mixing modulation $H_\textrm{mix1}=\Omega_\textrm{CP}(t)\hat{J}_{x}$ (implemented by the CP sequences) is applied to obtain the phase accumulation $\phi^\textrm{CP}_n$.
Here, we set the CP sequences and PDD sequences such that $\Omega(t)=\Omega$ when $\textrm{mod}[n,4]=1,2$, and $\Omega(t)=-\Omega$ when $\textrm{mod}[n,4]=3,0$, see Fig.~\ref{Fig1}.
The CP sequences and PDD sequences we used can resist the rotation angle and the detuning error(see Sec.~\ref{Sec41}
for more details).
In the readout stage, a certain unitary operation $\hat{U}_{2}$ is applied for recombination, and the half-population difference is measured.
The choices of unitary operations depend on the input state and have influences on the final measurement precisions.
The final state before half-population difference measurement can be written as
\begin{eqnarray}\label{Squential_scheme_final_state}	 \ket{\psi}_{\textrm{f}}=\hat{U}_{2}e^{-in\pi\hat{J}_x}e^{-i\phi_n^\textrm{CP}\hat{J}_{z}}\hat{U}_{1}e^{-i(n-1)\pi\hat{J}_x}e^{-i\phi_n^\textrm{PDD}\hat{J}_{z}}\ket{\psi}_{\textrm{in}}\nonumber\\
\end{eqnarray}
%
with $\phi^\textrm{PDD}_n$ is the accumulated phases via PDD sequences and it is
\begin{eqnarray}\label{phi_PDD}
\phi^\textrm{PDD}_n&=&\frac{2A}{\omega}\cos\left[\frac{n\omega\cdot(\tau_m-\tau)}{2}+\beta\right]\\\nonumber
&\times& \cos\left[\frac{\omega\cdot(\tau_m-\tau)}{2}\right]\frac{\sin\left[n\omega\cdot(\tau_m-\tau)/2\right]}{\sin\left[\omega\cdot(\tau_m-\tau)/2\right]},
\end{eqnarray}
and $\phi^\textrm{CP}_n$ is the accumulated phase via CP sequences and it is
\begin{eqnarray}\label{phi_CP}
\phi^\textrm{CP}_n&=&\frac{2A}{\omega}\sin\left[\frac{n\omega\cdot(\tau_m-\tau)}{2}+\beta+n\omega\tau_m\right]\\\nonumber
&\times&\left[1+\sin\left(\frac{\omega\cdot(\tau_m-\tau)}{2}\right)\right]\frac{\sin[n\omega\cdot(\tau_m-\tau)/2]}{\sin[\omega\cdot(\tau_m-\tau)/2]},
\end{eqnarray}
here $\tau={\pi}/{\omega}$ is the half period of the target signal.
When $\tau_m\neq\tau$, the two accumulated phase oscillates dramatically and are sensitively dependent on $\tau_m-\tau$.
While for $\tau_m=\tau$, the two accumulated phases are $\phi_n^\textrm{PDD}=\frac{2 A}{\omega}\cos(\beta)\cdot n$ and $\phi_n^\textrm{CP}=\frac{2 A}{\omega}\sin(\beta)\cdot n$ respectively.
%
%
The expectation of the half-population difference measurement on the final state is
\begin{eqnarray}\label{Jz_SCS2_lock_in}
\langle J_{z} \rangle_{\text{f}}=\bra{\psi}_{\textrm{f}}\hat{J}_{z}\ket{\psi}_{\textrm{f}}
\end{eqnarray}
%
According to quantum estimation theory, the measurement precision of the estimated parameters is given by the error propagation formula~\cite{Helstrom,Helstrom1967,Paris2009}.
\begin{equation}\label{Eq:Parameter uncertainty}
\Delta \mu=\frac{(\Delta{\hat{J}_{z}})_{\text{f}}}{|\partial{\langle\hat{J}_{z}\rangle_{\text{f}}}/ \partial{\mu}|},\quad (\mu=A,\omega,\beta)
\end{equation}
with $(\Delta{\hat{J}_{z}})_{\text{f}}$ is the standard deviation in the form of
\begin{equation}\label{Eq:Deviation}
(\Delta{\hat{J}_{z}})_{\text{f}}=\sqrt{\langle\hat{J}_z^2\rangle_{\text{f}}-(\langle{\hat{J}_{z}\rangle_{\text{f}}})^2},
\end{equation}
In the following, we investigate the measurement precisions with individual and entangled particles within our scheme.
\begin{figure*}[!htp]
\includegraphics[width=2\columnwidth]{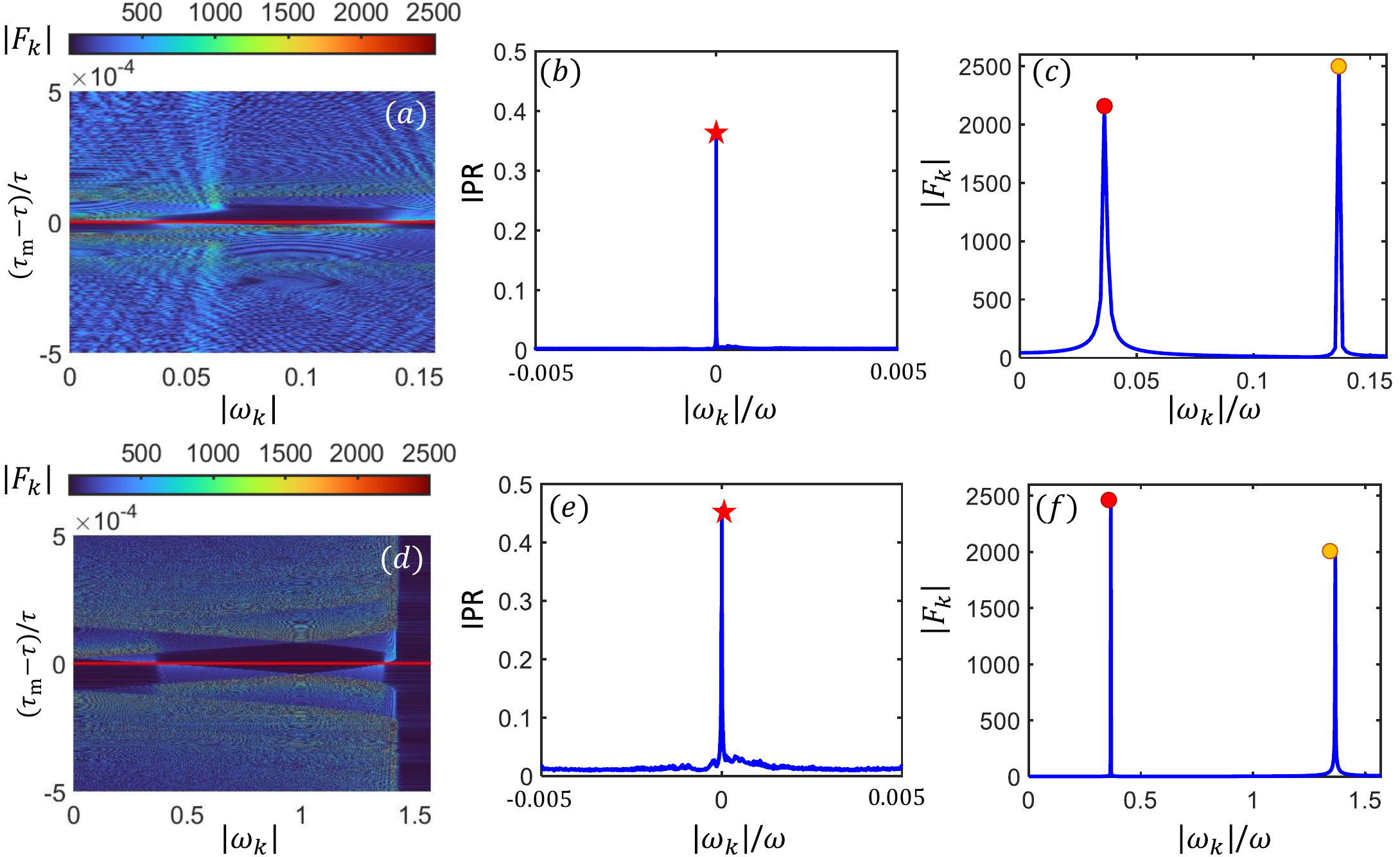}
\caption{\label{Fig2}
 Extraction of target signal with our scheme for individual and entangled particles.
 The fast Fourier transform (FFT) spectrum of the measurement signal $\langle\hat{J}_{z}\rangle_{\text{f}}$ versus $(\tau_m-\tau)$ with even positive integers $n$ up to $n_m=1000$ for (a)SCS and (d)GHZ state.
The inverse participation ratio (IPR) versus $\tau-\tau_m$ for (b)SCS and (e)GHZ state.
Given $\tau-\tau_m$, the FFT spectrum of the measurement signal $\langle\hat{J}_{z}\rangle_{\text{f}}$ at the lock-in point for (c)SCS and (f)GHZ state, the spectrum just has two peaks and one can use it to determine the lock-in point.
The two peaks locate at $\omega_1^\textrm{SCS}/\omega=|2A[\sin(\beta)-\cos(\beta)]|$,  $\omega_2^\textrm{SCS}/\omega=|2A[\sin(\beta)+\cos(\beta)]|$ for SCS state.
The two peaks are 0.0361 and 0.1366 which are very close to the theoretical $\omega_1^\textrm{SCS}/\omega=0.0366$,  $\omega_2^\textrm{SCS}/\omega=0.1366$.
The two peaks locate at $\omega_1^\textrm{GHZ}/\omega=|2NA[\sin(\beta)-\cos(\beta)]|$,  $\omega_2^\textrm{GHZ}/\omega=|2NA[\sin(\beta)+\cos(\beta)]|$ for GHZ state.
The two peaks are 0.3659 and 1.366 which are very close to the theoretical $\omega_1^\textrm{GHZ}/\omega=0.366$,  $\omega_2^\textrm{GHZ}/\omega=1.366$.
Here, we choose $A=\pi/2$, $\beta=\pi/6$, $\omega=10\pi$ and $N=10$.
}
\end{figure*}
\section{Measurement precisions\label{Sec3}}
In the following, we illustrate how to estimate the three parameters and give the measurement precisions with individual and entangled particles.
For individual particles without entanglement, the measurement precisions for the three parameters can only approach SQL.
For entangled particles in GHZ state, the measurement precision for the three parameters can both exhibit  Heisenberg scaling by using two suitable
interaction-based operations.
\subsection{INDIVIDUAL PARTICLES\label{A}}

First we consider the scheme with individual particles and assume the probe is prepared in the spin coherent state (SCS) $|\textrm{SCS}\rangle=e^{i\frac{\pi}{2}\hat J_y}\ket{N/2,-N/2}$.
This input state can be easily generated by applying a $\pi/2$ pulse on the state of all particles in spin-down $\ket{\downarrow}$.
Under this situation, one can choose $\hat{U}_{1}=\hat{U}_{2}=e^{-i\frac{\pi}{2}\hat{J}_x}$.
Then, the final state before measuring the half-population difference is
\begin{eqnarray}\label{Evo_CSC}
|\psi\rangle_{\text{f}}\!=\!e^{\!-i\frac{\pi}{2}\!{\hat{J}_{x}}}e^{\!-i\phi_n^\textrm{CP}\hat{J}_z}e^{\!-i\frac{\pi}{2}\!{\hat{J}_{x}}} e^{\!-i(n-1)\pi\!\hat{J}_x}e^{\!-i\phi_n^\textrm{PDD}\hat{J}_z}\!\ket{\textrm{SCS}}.\nonumber\\
\end{eqnarray}
After some algebra, the expectations of half-population difference on the final state is
\begin{eqnarray}\label{Jz_SCS2}
\langle J_{z} \rangle_{\text{f}}&=&\frac{N}{2}\sin(\phi_n^\textrm{CP})\cos(\phi_n^\textrm{PDD})\\\nonumber
&=&\frac{N}{4}[\sin(\phi_n^\textrm{CP}\!-\!\phi_n^\textrm{PDD})\!+\!\sin(\phi_n^\textrm{CP}\!+\!\phi_n^\textrm{PDD})].
\end{eqnarray}
At the lock-in point $\tau_m=\tau$, it reads
\begin{eqnarray}\label{Jz_SCS2_lock_in}
\langle J_{z} \rangle_{\text{f}}&=&\frac{N}{4}\sin\left[\frac{2 A}{\omega}\left(\sin(\beta)-\cos(\beta)\right)\cdot n\right]\\\nonumber
& &-\frac{N}{4}\sin\left[\frac{2 A}{\omega}\left(\sin(\beta)+\cos(\beta)\right)\cdot n\right]
\end{eqnarray}
which is an exactly bisinusoidal oscillation.
In practical quantum sensors, the fast Fourier transform  (FFT) method has been widely used in magnetic field measurement.
Similar to the method used in Ref~\cite{Chen2024}, one can determine the lock-in point via FFT by checking whether the measurement signal $\langle J_{z} \rangle_{\text{f}}$ shows a bisinusoidal oscillatory pattern.
%
As shown in Fig.~\ref{Fig2}~(a), we give the FFT spectrum of $\langle J_{z} \rangle_{\text{f}}$ versus $\tau-\tau_m$.
When $\tau=\tau_m$, the FFT spectrum has only two peaks and therefore one also can determine the lock-in point via its inverse participation ratio (IPR)~\cite{NCMurphy2011,MCalixto2015,FEvers2000,TBPClark2018,MYamanaka2018}, which is defined as
\begin{equation}\label{IPR}
\textrm{IPR}=\frac{\sum_{k=1}^{n_m/2}|F_k|^4}{\left|\sum_{k=1}^{n_m/2}|F_k|^2\right|^2},
\end{equation}
$|F_k|$ is the FFT amplitude corresponding to FFT angular frequency $\omega_{k}$.
After some algebra, we have ${\textrm{IPR}={1}/{2}}$ for $\tau=\tau_m$ and ${\textrm{IPR}=0}$ for $\tau\neq\tau_m$.
%
As shown in Fig.~\ref{Fig2}~(b), we have $\textrm{IPR}\approx 1/2$ for $\tau_m=\tau$ and $\textrm{IPR}\approx0$ for $\tau_m\neq\tau$, thus one can extract the frequency $\omega$ by finding the maximum of $\textrm{IPR}$.
%
%
At the lock-in point, the two peaks appear at $\omega_1^\textrm{SCS}=|2A[\sin(\beta)-\cos(\beta)]|$,  $\omega_2^\textrm{SCS}=|2A[\sin(\beta)+\cos(\beta)]|$, thus the values of $A$ and $\beta$ are given as
\begin{eqnarray}\label{A}
A=\frac{1}{2\sqrt{2}}\sqrt{\left(\omega_1^\textrm{SCS}\right)^2+\left(\omega_2^\textrm{SCS}\right)^2}
\end{eqnarray}
and
\begin{eqnarray}\label{S_beta_GHZ}
\beta=\frac{1}{2}\arcsin\left[\frac{\left(\omega_2^\textrm{SCS}\right)^2-\left(\omega_1^\textrm{SCS}\right)^2}{8A^2}\right]
\end{eqnarray}
%
%
The FFT spectrum for $N=10$ at the lock-in point is shown in Fig.~\ref{Fig2}~(c), the numerical results are well agree with our theoretical predictions.
Moreover, one can determine the sign of parameters $\beta$ from $\langle \hat J_{z} \rangle_{\text{f}}$ via a fitting procedure. 

Now we study the measurement precision for $\omega$, $A$, $\beta$ with individual particles.
After some algebra, the expectations of the square of the half-population difference is 
\begin{eqnarray}\label{Jz2_SCS2}
\langle J_{z}^2 \rangle_{\text{f}}=\frac{N}{4}+\frac{N(N-1)}{4}\sin(\phi_n^\textrm{CP})^2\cos(\phi_n^\textrm{PDD})^2.
\end{eqnarray}
According to Eq.~\eqref{Eq:Parameter uncertainty}, we can analytically obtain $\Delta\mu (\mu=\omega, A, \beta)$ and they read as
\begin{eqnarray}\label{S_Delta_mu_SCS}
\Delta \mu=\frac{1}{\sqrt{N}}\frac{{\sqrt{1-\left[\sin(\phi_n^\textrm{CP})\cos(\phi_n^\textrm{PDD})\right]^2}}}{
|G|}.\nonumber\\
\end{eqnarray}
with 
\begin{eqnarray}\label{g}
G\!=\!\partial_\mu\phi_n^\textrm{PDD}\!\!\sin(\phi_n^\textrm{PDD})\!\sin(\phi_n^\textrm{CP})
\!-\!\partial_\mu\phi_n^\textrm{CP}\!\!\cos(\phi_n^\textrm{CP})\!\cos(\phi_n^\textrm{PDD})\nonumber\\
\end{eqnarray}
%
%
According to Eq.~\eqref{S_Delta_mu_SCS}, it is obvious that the measurement precisions $\Delta\mu$ follows the SQL scaling (i.e., $\Delta \mu\propto1/\sqrt{N}$)
As shown in {Fig}.~\ref{Fig3}, we numerically find the measurement precisions $\Delta \mu$ versus the total particle number $N$.
According to the fitting results, the log-log measurement precisions for $\textrm{ln}\left(\Delta A\right)\approx -0.5\textrm{ln}({N})-5.4$ (red triangle), $\textrm{ln}\left(\Delta\omega\right)\approx-0.5\textrm{ln}({N})-2.4$ (red triangle) and $\textrm{ln}\left(\Delta\beta\right)\approx -0.5\textrm{ln}({N})-3$ (red triangle).
Since the input state is not entangled, the measurement precisions for the three parameters just can saturate the SQL.
\begin{figure*}[!htp]
\includegraphics[width=2\columnwidth]{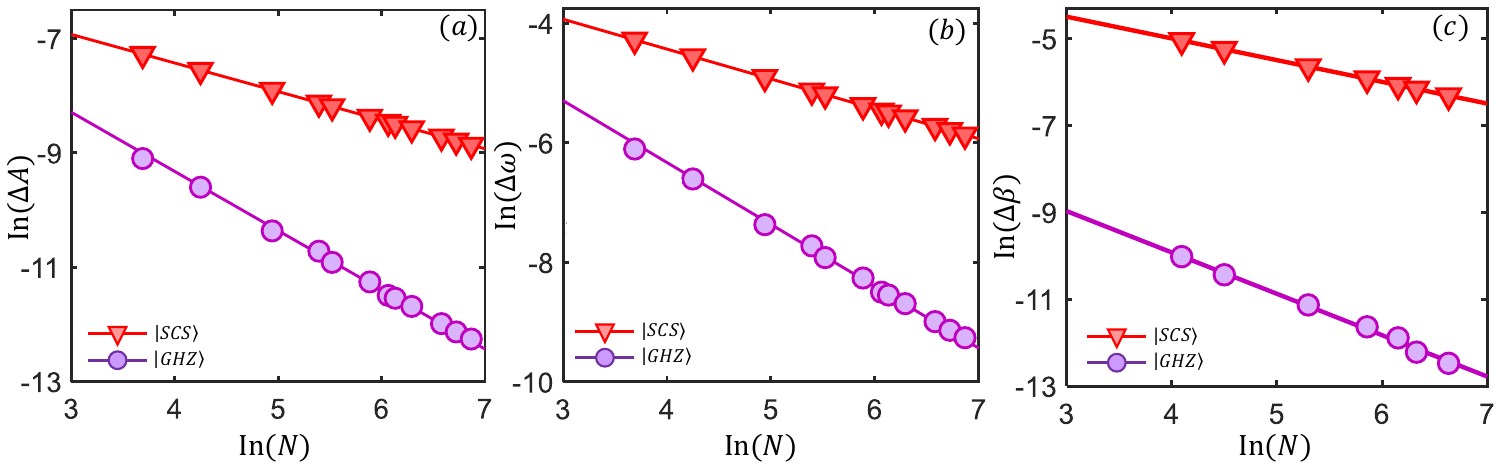}
\caption{\label{Fig3}
Log-log scaling of measurement precisions (a)$\Delta{A}$,(b)$\Delta\omega$, (c) $\Delta\beta$ with respect to the total particle number $N$ for our scheme.
The triangles correspond to the SCS state, and the circles correspond to the GHZ state.
Here $A=\pi/2$, $\beta=\pi/6$, and $\omega=10\pi$.}
\end{figure*}

\subsection{ENTANGLED PARTICLES\label{B}}

In this section, we discuss the scheme with entangled particles and show how to realize the Heisenberg-limited measurement of the target signal.
Entanglement is a useful quantum resource to improve the measurement precision.
%
%
Here, we choose the GHZ state $|\textrm{GHZ}\rangle=(\ket{N/2,N/2}+\ket{N/2,-N/2})/\sqrt{2}$ as an input state.
%
Meanwhile, it is known that interaction-based readout is a powerful technique for achieving the Heisenberg limit via a GHZ state without single-particle-resolved detection~\cite{PRApplied13044049,PRL116053601,PRL116090801,PRA98012129,PRL119193601,Mirkhalaf2018},
and is now feasible in experiments~\cite{Science3641163,PRL117013001}.
In our scheme, we choose the interaction-based operation
$\hat{U}_{1}=\hat{U}_{2}=e^{-i\frac{\pi}{2}{\hat{J}_{y}}}e^{-i\frac{\pi}{2}{\hat{J}_{z}^2}}e^{i\frac{\pi}{2}{\hat{J}_{y}}}$.
After the quantum interferometry, the final states before the half-population difference measurement can be written as
\begin{eqnarray}\label{Evo_Spin_cat_state3}
\ket{\psi}_{\text{f}}&\!=\!& e^{-i\frac{\pi}{2}\hat{J}_y}e^{-i\frac{\pi}{2}\hat{J}_z^2}e^{i\frac{\pi}{2}\hat{J}_y} e^{\!-in\pi\!\hat{J}_x} e^{-i\phi_n^\textrm{CP}\hat{J}_z}
\nonumber \\
&\!\times\!& e^{-i\frac{\pi}{2}\hat{J}_y}e^{-i\frac{\pi}{2}\hat{J}_z^2}e^{i\frac{\pi}{2}\hat{J}_y} e^{\!-i(n-1)\pi\!\hat{J}_x} e^{-i\phi_n^\textrm{PDD}\hat{J}_z}
\ket{\textrm{GHZ}}.\nonumber \\
\end{eqnarray}
After some algebra, the final expectations of half-population is
\begin{eqnarray}\label{Jz_GHZ}
\langle J_{z} \rangle_{\text{f}}=\frac{(-1)^{\frac{N+2}{2}}\!N}{4}\!
\sin[N(\phi_n^\textrm{CP}-\phi_n^\textrm{PDD})]\\
+\frac{(-1)^{\frac{N+2}{2}}\!N}{4}\!\sin[N(\phi_n^\textrm{CP}+\phi_n^\textrm{PDD})]\nonumber \\
\end{eqnarray}
Because of the entanglement, the oscillation of the measurement signal $\langle J_{z} \rangle_{\text{f}}$ becomes related to $N$.
At the lock-in point $\tau_m=\tau$, it reads
\begin{eqnarray}\label{JzGHZ_lock_in}
\langle J_{z} \rangle_{\text{f}}&=&\frac{N}{4}\sin\left[\frac{2N A}{\omega}\left(\sin(\beta)-\cos(\beta)\right)\cdot n\right]\\\nonumber
& &-\frac{N}{4}\sin\left[\frac{2N A}{\omega}\left(\sin(\beta)+\cos(\beta)\right)\cdot n\right]
\end{eqnarray}
The Eq.~\eqref{JzGHZ_lock_in} is also an exactly bisinusoidal oscillation, thus one also can obtain the lock-in point via the FFT.
As shown in Fig.~\ref{Fig2}~(d), we give the FFT spectrum of $\langle J_{z} \rangle_{\text{f}}$ versus $\tau-\tau_m$.
When $\tau=\tau_m$, the FFT spectrum exhibits only two peaks, thus one can also determine the lock-in point via its IPR.
Similarly,  we have ${\textrm{IPR}={1}/{2}}$ for $\tau_m=\tau$ and ${\textrm{IPR}=0}$ for $\tau_e\neq\tau$, as shown in Fig.~\ref{Fig2}(e).
At the lock-in point, the two peaks appear at $\omega_1^\textrm{GHZ}=|2NA[\sin(\beta)-\cos(\beta)]|$,  $\omega_2^\textrm{GHZ}=|2NA[\sin(\beta)+\cos(\beta)]|$.
Thus the values of $A$ and $\beta$ are given as
\begin{eqnarray}\label{A}
A=\frac{1}{2\sqrt{2}N}\sqrt{\left(\omega_1^\textrm{GHZ}\right)^2+\left(\omega_2^\textrm{GHZ}\right)^2}
\end{eqnarray}
and
\begin{eqnarray}\label{S_beta_GHZ}
\beta=\frac{1}{2}\arcsin\left[\frac{\left(\omega_2^\textrm{GHZ}\right)^2-\left(\omega_1^\textrm{GHZ}\right)^2}{8N^2A^2}\right]\nonumber\\
\end{eqnarray}
The FFT spectra for $N = 10$ at the lock-in point is shown in Fig.~\ref{Fig5}(f).
The numerical results perfectly agree with our theoretical predictions.
Meanwhile, the square of the half-population on the final state is $\langle J_{z}^2 \rangle_{\text{f}}=\frac{N^2}{4}$.
According to Eq.~\eqref{Eq:Parameter uncertainty}, we analytically obtain $\Delta \mu$ and they are
\begin{eqnarray}\label{Delta_mu_GHZ}
\Delta \mu=\frac{1}{N}\frac{\sqrt{1-[\sin(N\phi_n^\textrm{CP})\cos(N\phi_n^\textrm{PDD})]^2}}{|G|}.
\end{eqnarray}
According to Eq.~\eqref{Delta_mu_GHZ}, the measurement precisions $\Delta \mu$ for individual particles can exhibit the HL scaling(i.e.,$\Delta \mu \propto 1/{N}$).
%
%
%
As shown in Fig.~\ref{Fig3}, according to the fitting results, the log-log measurement precisions for $\textrm{ln}\left(\Delta A\right)\approx -\textrm{ln}({N})-5.2$ (purple circle), $\textrm{ln}\left(\Delta\omega\right)\approx-\textrm{ln}({N})-2.2$ (purple circle) and $\textrm{ln}\left(\Delta\beta\right)\approx -0.95\textrm{ln}({N})-6.1$ (purple circle).

\section{Robustness against imperfections}\label{Sec4}
In this section, we study the robustness of our scheme.
In practical experiments, there are many imperfections
that can limit the final measurement precision.
Here, we discuss three
imperfections: the rotation angle errors and detuning errors in the two phase accumulation processes, and the detection noise in the measurement
step.
\begin{figure*}[!htp]
\includegraphics[width=2\columnwidth]{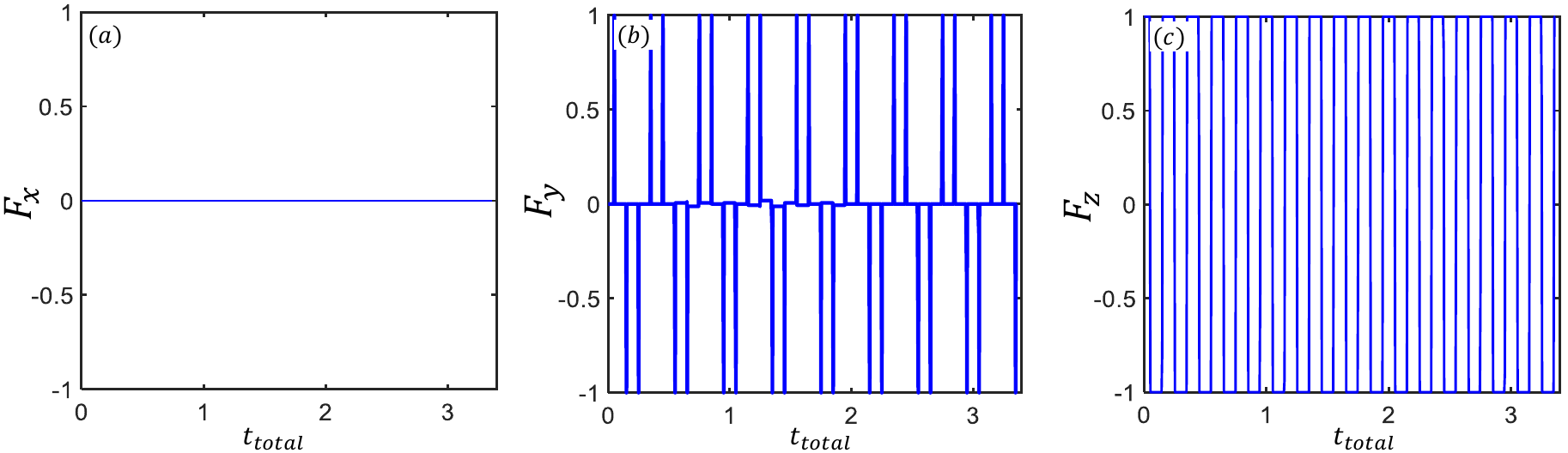}
\caption{\label{Fig3}
	 Time evolution of the coefficients  (a) $F_x$, (b)$F_y$, and (c) $F_z$ for our scheme.
     This indicates that Eq.~\eqref{Conditionerror} and Eq.~\eqref{Conditiondetuning} are both satisfied in our scheme}
\end{figure*}

\subsection{Robustness against rotation angle errors and detuning errors}\label{Sec41}
To realize perfect interrogation in our schemes, the
pulses should be ideal $\pi$ pulses. 
However, in practical experiments, laser intensity fluctuations and frequency instability are the main noise sources, which induce rotation angle errors and detuning errors respectively.
Here, to illustrate the robustness of our scheme, we make
use of the transformations of the signal terms in the interaction picture by executing periodic pulse sequences under these noises.
Thus, we consider in the interaction picture the signal term ${J}_z$ under a given periodic pulse sequence $[P_k;k=1,2,...n]$ with each pulse of the same form of $P_k=e^{-{i\Omega(t)\tau_{\pi}}\hat{J}_{z}}$ (with $\tau_{\pi}$ being the $\pi$ pulse length).
The transformation trajectory in the toggling frame can be identified as
\begin{eqnarray}\label{Conditiondetuning1}
\widetilde{J}_{z}&=&(P_{k-1}\cdots P_{1})^{\dag}\hat{J}_{z}(P_{k-1}\cdots P_{1})\nonumber\\
&=&\int_{(k-1)\tau}^{k\tau}  F_{\mu}(t)\hat{J}_{\mu}dt~~~(\mu=x,y,z),
\end{eqnarray}
with $\tau$ being the pulse spacing between two adjacent pulses.
Intuitively, a nonzero element ${F}_{\mu}(t)$ indicates that the initial $\hat{J}_{z}$ operator transforms into $\hat{J}_{\mu}$ during the free-evolution interval $\tau$.
Based on this representation, we can formulate the conditions for suppressing the rotation angle and detuning errors.
Theoretical illustration and extension of this framework, as well as the use for other applications and Hamiltonians, can be found in Refs. \cite{Choi2020,Zhou2024,Choi2017}.
The condition for suppressing the detuning error is
\begin{eqnarray}\label{Conditiondetuning}
	\int_{0}^{t_\textrm{total}}{F}_{\mu}(t)dt=0 ~~~(\mu=x,y,z)
\end{eqnarray}
with $t_\textrm{total}=n(\tau_\pi+\tau)$ denoting the total evolution time.
The intuition behind the rule for suppressing the detuning error is simply that a spin echo must be performed along each axis direction, so that a precession around a positive detuning axis (e.g., $+\hat{J}_z$) is compensated by a negative precession (e.g., $-\hat{J}_z$).
Thus it requires equal amounts of the evolution time along the positive and negative directions for each axis and can be described via Eq.~\eqref{Conditiondetuning1}.

The condition for suppressing the rotation angle errors is
\begin{eqnarray}\label{Conditionerror}
\int_{0}^{t_\textrm{total}}[\textbf{{F}(t)} \times \textbf{{F}(t)}]dt=\vec{0},
\end{eqnarray}
where $\textbf{{F}(t)}=\sum_{\mu}{F}_{\mu}(t)\hat{e}_{\mu}$ with $\hat{e}_{\mu}$ being the $\mu$-axis unit vector.
The physics behind Eq.~\eqref{Conditionerror} is that a systematic rotation along the $+\hat{\mu}$-axis (positive chirality) can be compensated by another rotation along the $-\hat{\mu}$-axis (negative chirality) in the toggling frame.
This chirality can be conveniently described mathematically by the cross product between the neighboring frame directions, yielding a simple algebraic description, which gives the simple algebraic condition Eq.~\eqref{Conditionerror} for counteracting rotation angle errors.

In our consideration, the total evolution time is $t_{total}=n(\tau+\tau_{\pi})$.
As shown in Fig.~\ref{Fig3}, the coefficients  ${F}_{x}$, ${F}_{y}$ and ${F}_{z}$ occupy equal areas with respect to the zero point of the vertical axis in each of the three panels.
This indicates that both Eq.~\eqref{Conditionerror} and Eq.~\eqref{Conditiondetuning} are well satisfied, and thus our scheme can resist the rotation angle and detuning errors.
\begin{figure*}[!htp]
	\includegraphics[width=2\columnwidth]{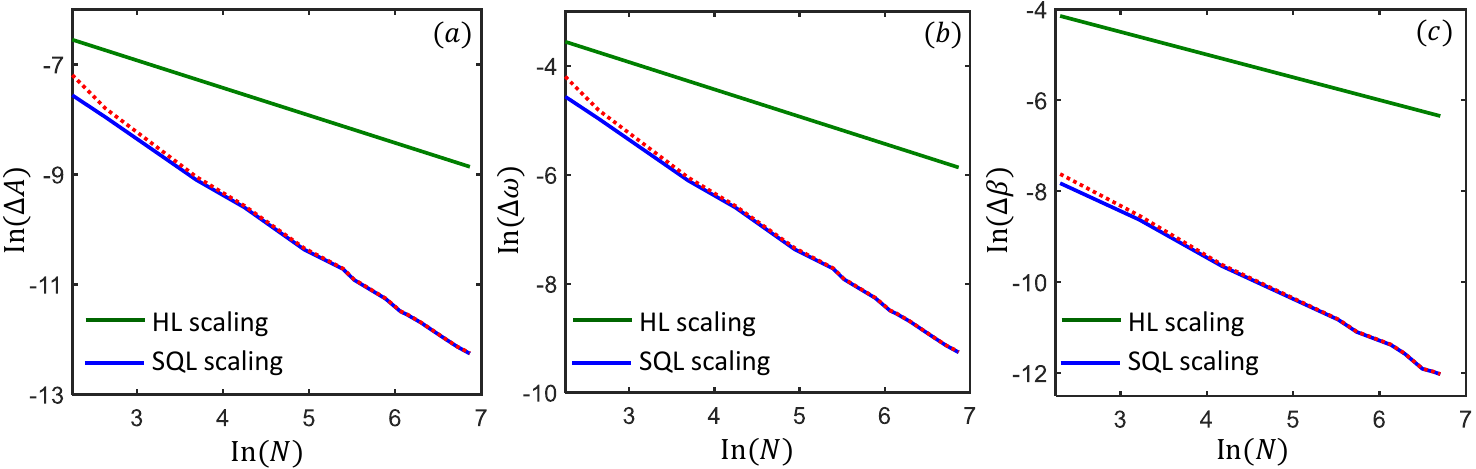}
	\caption{\label{Fig5}(color online).
		Log-log scaling of measurement precisions (a)$\Delta{A}$,(b)$\Delta\omega$,(c) $\Delta\beta$ with respect to the total particle number $N$ for sequential scheme with GHZ state under detection noise $\sigma$.
  Here $A=\pi/2$, $\beta=\pi/6$, and $\omega=10\pi$, $\sigma=0.5\sqrt{N}$ (purple dotted line), which are somewhere between SQL (blue dashed line) and HL (black solid line)
	}
\end{figure*}
\subsection{Robustness against detection noise}\label{Sec42}
Here, we study the influence of detection noise by considering
the additional classical noise in the measurement process.
%
%
%
%
%
%
%
Considering the Gaussian detection noise $\sigma$, the variance of the population difference becomes $(\Delta\tilde{J}_z)^2=(\Delta J_z)^2+\sigma^2$~\cite{Nolan2017,PRApplied064056,JHarXiv2022}, thus the measurement precision  will also degrade.
Our results show that the detection noise with $\sigma=\sqrt{N}$ indeed will also limit precision scaling to somewhere between SQL and HL, as shown in Fig.~\ref{Fig5}.
The influence of detection noise can be neglected when the partical number is large. 

\section{SUMMARY\label{Sec5}}
We have presented a general scheme for realizing a quantum double lock-in detection via a single quantum interferometry with sequential orthogonal periodic multipulse sequences.
%
In our scheme, the interrogation stage is divided into two signal accumulation processes. 
In the first accumulation process, the PDD sequences is applied to realize one quantum mixing.
In the second accumulation process, the CP sequences is applied to realize other quantum mixing.
Additionally, the XY4-N sequences can also be used to realize the two orthogonal periodic multi-pulse sequence.
On the basis of our scheme, the complete information about the signal’s amplitude, frequency, and initial phase can be extracted from the FFT spectra of the half-population.
%
%
%
Compared to the conventional double lock-in detection with two individual quantum interferometry, our scheme can reduce experimental resources.

Based upon the proposed scheme, we study the measurement precision of the target signal with individual and entangled particles.
%
%
Notably, if a GHZ state is employed as the input state and two suitable interaction-based operations are applied,  the measurement precisions of both singnal’s amplitude, frequency, and initial phase can exhibit Heisenberg-limited scaling. 
Moreover, the robustness of our scheme is also discussed.
Based on state-of-the-art techniques, our study not only provides a effective pathway to achieving Heisenberg-limited detection of an oscillating singnal, but also beneficial for the development of practical entanglement-enhanced quantum technologies,  including magnetometers~\cite{RevModPhys90035005}, atomic clocks ~\cite{PRL793865,PRL125210503}, and weak-force detectors~\cite{NatCommun814157}.

\acknowledgments{Min Zhuang and Sijie Chen contribute equally. This work is supported by the National Natural Science Foundation of China (Grants No. 12025509, No12305022, and No. 92476201), the National Key Research and Development Program of China (Grant No. 2022YFA1404104), and the Guangdong Provincial Quantum Science Strategic Initiative (GDZX2305006 and GDZX2405002).}

\end{document}